\documentclass[aps,prd,preprint,tightenlines,nofootinbib]{revtex4}

\usepackage{amsmath}
\usepackage{graphicx}
\DeclareGraphicsExtensions{.ps}

\def\gtorder{\mathrel{\raise.3ex\hbox{$>$}\mkern-14mu
	\lower0.6ex\hbox{$\sim$}}}
\def\ltorder{\mathrel{\raise.3ex\hbox{$<$}\mkern-14mu
	\lower0.6ex\hbox{$\sim$}}}

\def \bea{\begin{eqnarray}}
\def \beq{\begin{equation}}
\def \eea{\end{eqnarray}}
\def \eeq{\end{equation}}

\begin{document}

\rightline{EFI 09-21}
\rightline{arXiv:1005.0797}

\centerline{\bf ELECTROWEAK CONSTRAINTS FROM}
\centerline{\bf ATOMIC PARITY VIOLATION AND NEUTRINO SCATTERING}
\bigskip

\centerline{\it Timothy Hobbs}
\centerline{Department of Physics, Indiana University, Bloomington, IN 47405}
\medskip
\centerline{and}
\medskip
\centerline{\it Jonathan L. Rosner}
\centerline{Enrico Fermi Institute and Department of Physics}
\centerline{University of Chicago, Chicago, IL 60637}
\medskip
\centerline{\today}
\medskip

\centerline{\bf Abstract}

\begin{quote}

Precision electroweak physics can provide fertile ground for uncovering
new physics beyond the Standard Model (SM).  One area in which new physics
can appear is in so-called ``oblique corrections," {\it i.e.}, next-to-leading
order expansions of bosonic propagators corresponding to vacuum polarization.
One may parametrize their effects in terms of quantities $S$ and $T$ that
discriminate between conservation and non-conservation of isospin.  This
provides a means of comparing the relative contributions of precision
electroweak experiments to constraints on new physics.  Given the prevalence
of strongly $T$-sensitive experiments, there is an acute need for further
constraints on $S$, such as provided by atomic parity-violating experiments on
heavy atoms.  We evaluate constraints on $S$ arising from recently improved
calculations in the Cs atom.  We show that the top quark mass $m_t$ provides
stringent constraints on $S$ within the context of the SM.  We
also consider the potential contributions of next-generation neutrino
scattering experiments to improved $(S,T)$ constraints.

\end{quote}

\leftline{PACS numbers:  12.15.Ji, 12.15.Mm, 13.15.+g, 14.60.Lm}

\section{INTRODUCTION}
\label{sec:intro}

The search for exotic physics not explicable in terms of the standard 
$SU(3) \times SU(2) \times U(1)$ model has been carried out on the complementary
frontiers of energy and precision.  Exploration on the precision frontier has
included a wide range of measurement in electroweak physics, seeking to reduce
experimental uncertainties of common inputs and other observables of the
SM.  Present and proposed versions of such experiments
promise unprecedented measurements of such basic SM quantities as the
electroweak mixing angle, providing needed tests of the SM and searches for
Beyond-Standard-Model (BSM) physics.

A formalism that broadly incorporates the effects of hypothetical new physics
can be achieved through the introduction of vacuum polarization loop
corrections to the vector boson propagators of the electroweak theory; an
example of this modification is illustrated in Fig.~\ref{fig:diag} for the
leading-order (LO) photon propagator in QED. Such corrections modify the
LO theory in a way whereby effects of exotic physics outside the SM can be
represented in terms of a small number of ``oblique corrections" (OCs)
affecting gauge boson propagators.  These are to be distinguished from more
direct, one-loop corrections at the level of the vertex in the electroweak
theory.

It is possible to express obliquely-corrected quantities in terms of parameters
$S$ and $T$ \cite{PT}, permitting one to compare the relative contributions of
various precision electroweak experiments to consraints on new physics.
Several experiments constrain $T$ well, but there is a scarcity of constraints
on $S$, such as those provided by atomic parity-violating experiments on heavy
atoms \cite{MR}.

We briefly review the $(S,T)$ formalism in Sec.\ II.  We then evaluate in
Sec.\ III constraints on $S$ arising from recently improved calculations in
the Cs atom.  We also notice that within the Standard Model, the top quark
mass $m_t$ provides stringent constraints on $S$ (Sec.\ IV).  In Sec.\ V we
consider the potential of next-generation neutrino scattering experiments for
improving $(S,T)$ constraints. We conclude in Sec.\ VI.

% This is Section II
\section{OBLIQUE CORRECTIONS
\label{sec:ocs}}

\begin{figure}[t]
\includegraphics[height=6cm]{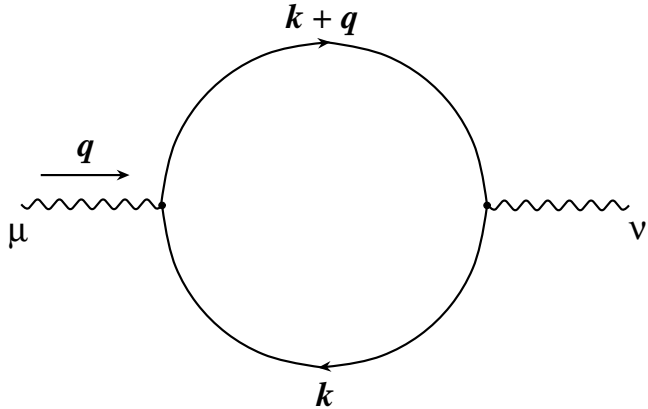}
\caption{The order-$\alpha$ correction to the photon propagator of QED.  This
diagram corresponds to the amplitude to spontaneously generate an
electron-positron pair from the vacuum; the mathematical details of this
process are similar in character to the self-energy corrections to
electroweak propagators.}
\label{fig:diag}
\end{figure}

The tools for approximating the effects of loop corrections in QCD are best
understood via analogy with the simpler $U(1)$ model of QED. In this setting,
the addition of a single loop correction associated with vacuum polarization
also modifies the photon propagator. So, the exchange of a photon with
4-momentum transfer $q^2$ is modified according to 
% Eq. (1)
\begin{equation}
\label{eq:qedm}
\frac{-i g_{\mu \nu}}{q^2} \rightarrow \frac{-i g_{\mu \nu}}{q^2} +
(\frac{-i g_{\mu \alpha}}{q^2}) \Pi^{\alpha \beta}
(\frac{-i g_{\beta \nu}}{q^2}),
\end{equation}
where the amplitude of the loop insertion is calculable according to the
Feynman diagram shown in Fig.~\ref{fig:diag}, viz.:
% Eq. (2)
\begin{equation}
\label{eq:qedl}
\Pi^{\alpha \beta}(q) = (-1) \int \frac{d^4k}{(2 \pi)^4} \text{Tr}
[(ie \gamma^\alpha) \frac{i({\not}k + m)}{k^2 - m^2} (ie \gamma^\beta)
\frac{i({\not}p + {\not}k + m)}{(q+k)^2 - m^2}] = (q^2 g^{\alpha \beta} -
q^\alpha q^\beta) \Pi(q^2)~.
\end{equation}
%
%This is Figure 1
As the integral in the loop amplitude is logarithmically divergent, it may be
renormalized by imposing a scale cutoff $\Lambda$ as
% Eq. (3)
\begin{equation}
\label{eq:intapp}
\Pi (q^2) \approx \frac{e_0^2}{12 \pi^2} \ln (\frac{\Lambda^2}{q^2})~.
\end{equation}
Thus, in this case, higher-order OCs may be recast as a
modification of the amplitude of the virtual photon exchange as an infinite
power series of logarithmically divergent integrals, with successive terms
suppressed by higher powers of $\alpha_0$. Standard inputs of the $U(1)$ theory
are then modified as
% Eq. (4)
\begin{equation}
\label{eq:alpha}
\alpha(q^2) = \alpha_0 (1 - \cdots \pm (\alpha_0 \Pi (q^2))^n) = 
\frac{\alpha_0}{1+\alpha_0 \Pi(q^2)}~.
\end{equation}
In the transition to the electroweak theory, the essential difference
from the foregoing QED ansatz for incorporating higher-order OCs
is the expanded dimensionality of the group of electroweak generators.
In an analogous fashion, the propagators of the vector bosons $Z$ and $W$
may also be expanded in terms of loop amplitudes proportional to logarithmically
divergent integrals. These amplitudes must be indexed over the SU(2) $\otimes$
U(1) group generators.  A particularly convenient means of collecting the
leading-order contributions of the indexed loop amplitudes is to write the
parameters \cite{PT}
\begin{subequations}
% Eq. (5a)
\begin{equation}
\label{eq:Sdef}
S = 16 \pi \frac{d}{dq^2}[\Pi_{33} (q^2) - \Pi_{3Q} (q^2)]_{q^2 = 0}
\end{equation}
and
% Eq. (5b)
\begin{equation}
\label{eq:Tdef}
T = \frac{4 \pi}{\sin^2 \theta_W \cos^2 \theta_W m_Z^2} 
[\Pi_{11} (q^2) - \Pi_{33} (q^2)]_{q^2 = 0}.
\end{equation}  
\end{subequations}
These parameters are suitable for computing and constraining OCs in that, for
the case of virtual $Z$-boson exchange, the modifications of the relevant
standard model inputs are expressible in terms of linear combinations of $S,
T$.  (We neglect here a parameter $U$ which arises only in particular models
and leads to a difference in $S$ when applied to $Z$ and $W$ propagators.)
One can thus represent SM observables for small deviations from nominal values
as $\bar{x} = \bar x_0 + A S + B T$, where $A, B$ are constants; $\bar x$ is a
measured quantity; and $\bar x_0$ is its value for nominal parameters
corresponding to $S=T=0$.  Such relations imply that measured values of SM
inputs, with associated experimental uncertainties, correspond to unique linear
bands in the $(S, T)$ plane.  The combination of bands with different slopes
constrains the space of OCs to an elliptical region.  In this way we may test
the effects on the allowed $(S,T)$ region of alterations in any experimental
inputs. For nominal inputs we take values quoted in Ref.\ \cite{Rosner:2001ck},
suitably updated with more recent calculations when available.
We use an effective value of the weak mixing angle as measured via leptonic
vector and axial-vector couplings:  $\sin^2 \theta^{\rm eff} \equiv (1/4)
(1 - [g^l_V/g^l_A])$. 

% This is Section III
\section{CONSTRAINTS FROM $Q_W({\rm Cs})$}

The orientation of the ellipse of OCs \cite{Rosner:2001ck}
suggests an acute need for additional constraints on $S$.  Measurements of
$Q_W$ in atomic parity violation experiments promise special sensitivity to 
$S$ with almost none to $T$ \cite{MR,Sandars}.  For this reason, there has been
great interest in improved measurements of $Q_W$ for heavy nuclei.  Aside from
atomic parity violation, the electroweak observables with the greatest ratio
of $S$ to $T$ coefficients (i.e., $A / B$) are the cross section for
$\bar \nu_e \to \bar \nu_e$, whose potential impact has been discussed in
Ref.\ \cite{Rosner:2004}, as well as the top quark mass $m_t$, whose effect
within the Standard Model is noted below in Sec.\ \ref{sec:top}.

The combination of measurements and theory provides the strongest constraints
at present for atomic cesium ($Z = 55$) \cite{Wood:1997,Guena:2005},
particularly in view of the latest theoretical atomic physics calculations
\cite{Porsev:2009pr}.  We therefore restrict our attention to cesium in what
follows.  For a cesium isotope of $N$ neutrons, the inclusion of
next-to-leading-order loop corrections yields \cite{MR}
% Eq. (6)
\begin{equation}
\label{eq:rmin1}
Q_W(_{55}^{55 + N} {\rm Cs}) = (0.9857 \pm 0.0004) \rho [-N + 55
(1 - (4.012 \pm 0.01) \sin^2 \theta_W)]~.
\end{equation}
If one assumes that custodial symmetry (corresponding to $\rho = 1$)
is broken, the previous expression may be expanded using $\rho = 1 + \alpha T$
and the standard, linearized expression $\sin^2 \theta^{\rm eff} = 0.2314 +
0.00361 S - 0.00257 T$ \cite{Rosner:2001ck}:
% Eq. (7)
\begin{equation}
\label{eq:rmin2}
Q_W(_{55}^{133} {\rm Cs}) = -73.16 - 0.800 S - 0.007 T~,
\end{equation}
in which we have taken the stable isotope with $N = 78$ for Cs.  A small change
from the central value of $-73.19$ quoted in Ref.\ \cite{Rosner:2001ck} is due
to corrections detailed in Ref.\ \cite{Erler,Amsler:2008}.  Clearly,
incremental reductions to the uncertainty of $Q_W$ will necessarily constrain
$S$ dramatically more tightly than $T$.  Updated calculations \cite{Erler,EL}
incorporating the atomic physics corrections of Ref.\ \cite{Porsev:2009pr} and
the two experimental measurements \cite{Wood:1997,Guena:2005} yield

% Eq. (8)
\beq
Q_W(_{55}^{133} {\rm Cs}) = -73.20(35)~.
\eeq
Neglecting the $T$ dependence in Eq.\ (\ref{eq:rmin2}), this implies a value
$S = 0.06 \pm 0.44$.  The error on $S$ has been reduced by more than a factor
of 2 since the analysis in Ref.\ \cite{Rosner:2001ck}.  We now discuss the
impact of this measurement in light of other electroweak constraints.

We use the same constraints as in an earlier analysis \cite{Rosner:2001ck},
suitably updated to reflect an erratum in Ref.\ \cite{NuTeV},
further corrections to the NuTeV result detailed in Ref.\ \cite{EL},
and the latest averages of $M_W$ and $\sin^2 \theta^{\rm eff}$ \cite{LEPEWWG}.
We omit an older constraint with large error bars based on parity violation in
atomic thallium.  The inputs are summarized in Table \ref{tab:inp}.

% This is Table I
\begin{table}
\caption{Electroweak observables leading to $S,T$ constraints.   
\label{tab:inp}}
\begin{center}
\begin{tabular}{l c c} \hline \hline
 & Experimental & Theoretical \\
Quantity & value & value \\ \hline
$Q_W({\rm Cs})$ & $-73.20 \pm 0.35$ & $-73.15 - 0.800S - 0.007T$ \\
$M_W({\rm GeV}/c^2)$ & $80.399 \pm 0.023$ & $80.385 - 0.29 S + 0.45 T$ \\
$g_L^2(\nu {\rm NC})$ \cite{NuTeV} & $0.3027 \pm 0.0018^{~a}$
 & $0.3041 -0.00272 S + 0.00665 T$ \\
$g_R^2(\nu {\rm NC})$ \cite{NuTeV} & $0.03076\pm 0.00110$ & $0.0300 + 0.00094 S
 - 0.00020 T$ \\
$\Gamma_{ll}(Z)$ (MeV) & $83.984^{~b} \pm 0.087$ & $84.005^{~b} - 0.18 S + 0.78
T$ \\
sin$^2 \theta^{\rm eff}$ & $0.23149 \pm 0.00016$ & $0.23140 + 0.00361 S
 - 0.00257 T$ \\ \hline \hline
\end{tabular}
\end{center}
\leftline{$^a$ Modification by \cite{EL} of value $g_L^2 = 0.30005 \pm 0.00137$
quoted in \cite{LEPEWWG}}
\leftline{$^b$ Values quoted in Ref.\ \cite{EL}}
\end{table}

\begin{figure}
\includegraphics[width=5in]{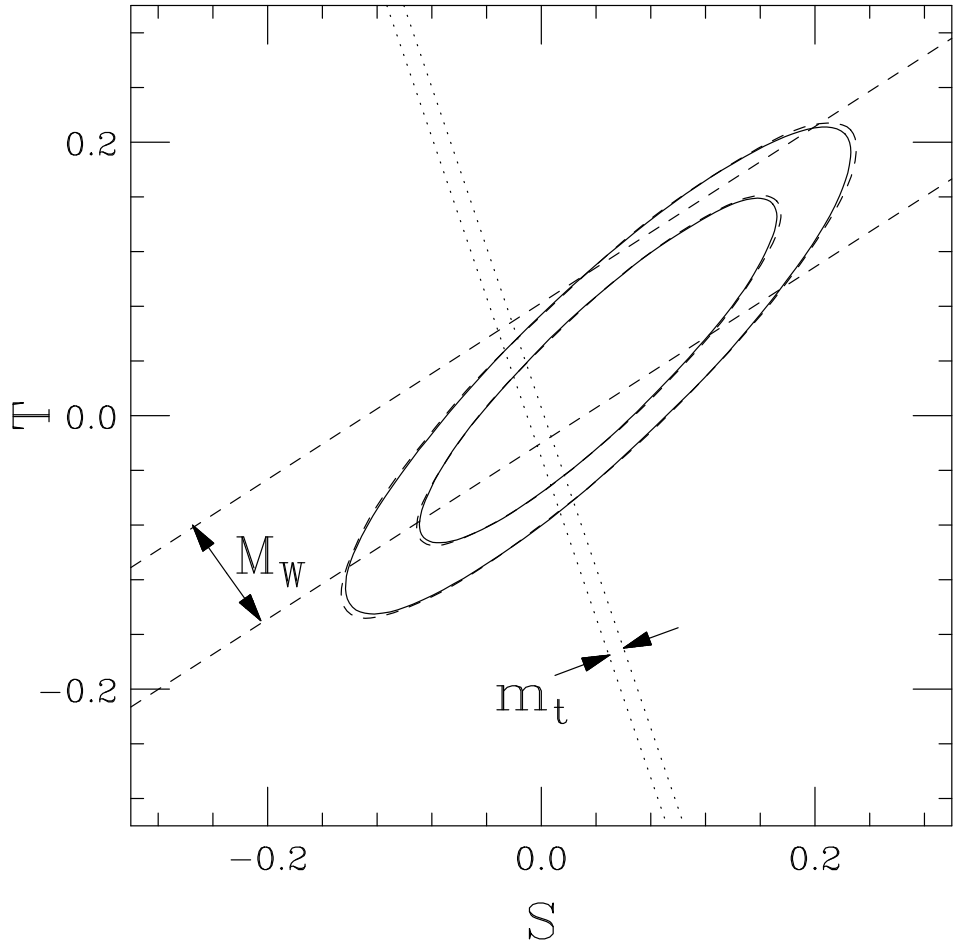}
\caption{Effects of atomic parity constraints in atomic Cs \cite{Wood:1997,%
Guena:2005,Porsev:2009pr}
on OCs in $S,T$ space. Inner and outer ellipses denote 68\% and
90\% confidence level (c.l.) limits.  Solid and dashed curves are plotted with
and without atomic parity constraints (first line of Table \ref{tab:inp}).
Dashed and dotted bands denote constraints due to $M_W$ (imposed) and $m_t$
(not imposed here but discussed below).
\label{fig:APV}}
\end{figure}

Constraints on $(S,T)$ due to the inputs in Table \ref{tab:inp} are summarized
in Fig.\ \ref{fig:APV}.  The (solid, dashed) ellipses correspond to constraints
(with, without) the atomic parity violation measurement on the first line
of Table \ref{tab:inp}.  The constraints due to $M_W$ (imposed) and $m_t$
(not imposed here, but to be discussed subsequently) are illustrated by the
dashed and dotted bands, respectively.

% This is Figure 2

Comparing the dashed and solid ellipses in Fig.\ \ref{fig:APV}, one sees that
the imposition of the constraints due to atomic parity violation has a small
effect on the allowed $(S,T)$ region.  Measurements at LEP and the
Fermilab Tevatron have provided sufficiently tight constraints that the value
of $S = 0.06 \pm 0.44$ from parity violation in atomic cesium is overshadowed
by the (approximate) constraint $|S|<0.16$ due to the remaining observables.
Thus, a further reduction in errors by a factor of at least three would be
required for atomic parity violation to significantly affect the constraints
on $(S,T)$.

% This is Section IV
\section{TOP QUARK MASS CONSTRAINTS
\label{sec:top}}

\begin{figure}
\includegraphics[width=5in]{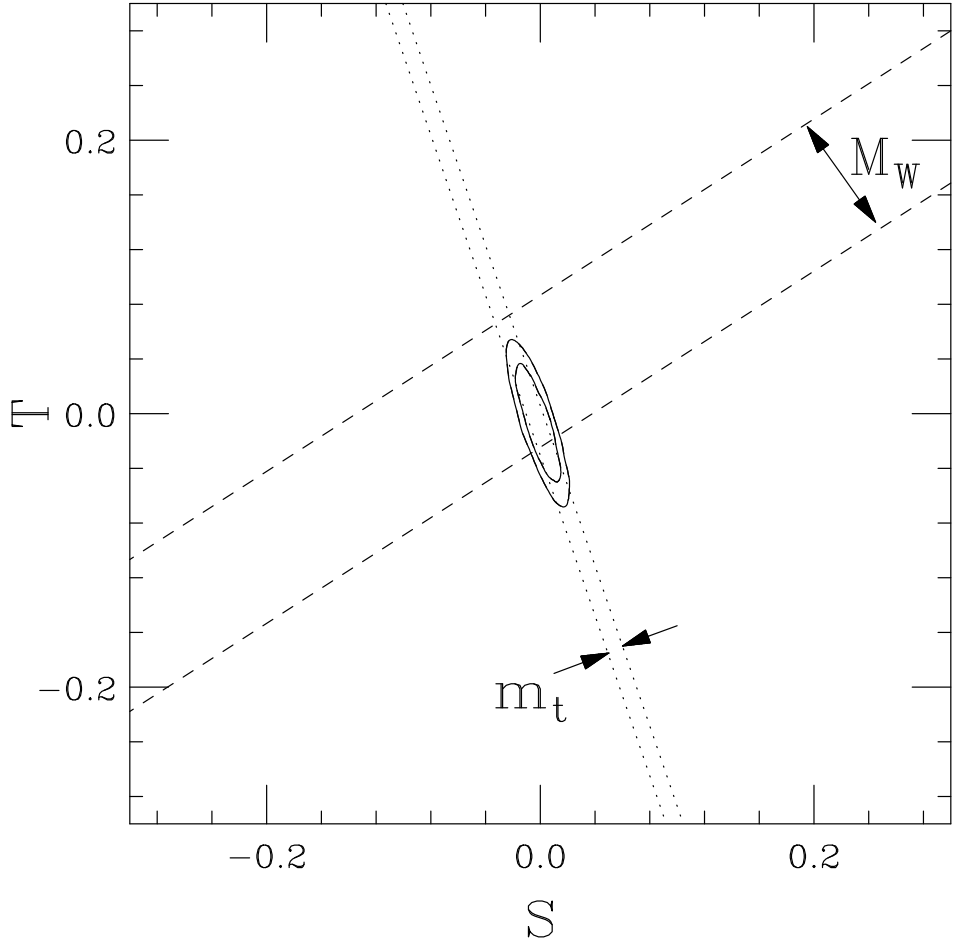}
\caption{Effects of including top quark mass constraint $m_t = 173.1 \pm 1.3$
GeV/$c^2$ on constraints in $S,T$ space.  Inner and outer ellipses denote 68\%
and 90\% confidence level (c.l.) limits.  Solid and dashed curves are plotted
with and without atomic parity constraints.  Dashed and dotted bands denote
constraints due to $M_W$ and $m_t$, respectively.
\label{fig:mt}}
\end{figure}

Instead of $(S,T)$ one often employs the measured quantities $(m_t, m_W)$ to
express constraints of electroweak measurements.  Using standard expressions to
approximate the logarithmic divergences of equations (\ref{eq:Sdef},
\ref{eq:Tdef}), $S$ and $T$ may be written \cite{MR}
% Eq. (9)
\begin{equation}
\label{eq:Sht}
S \approx \frac{1}{6 \pi} \ln \frac{m_H}{\Lambda_H}~,
\end{equation}
% Eq. (10)
\begin{equation}
\label{eq:Tht}
T \approx \frac{3}{16 \pi \sin^2 \theta_W} [\frac{m_t^2 - \Lambda_t^2}{m_W^2}]
- \frac{3}{8 \pi \cos^2 \theta} \ln \frac{m_H}{\Lambda_H}~.
\end{equation}  
Here we take the nominal parameters corresponding to $S=T=0$ to be $\Lambda_t =
174$ GeV (as in Ref.\ \cite{Rosner:2001ck}) and $\Lambda_H = 100$ GeV.  We may
then write
% Eq. (11)
\begin{equation}
\label{eq:TmtS}
T = \frac{3(m_t^2 - \Lambda_t^2)}{16 \pi \sin^2 \theta_W m_W^2} - \frac{9 S}
{4 \cos^2 \theta_W}~,
\end{equation}  
yielding a linear relation for the SM observable $m_t^2$ in terms of $S, T$:
% Eq. (12)
\begin{equation}
\label{eq:mtST}
m_t^2 = m_{W, 0}^2 [12 \pi \tan^2 \theta_W S + \frac{16 \pi \sin^2
\theta_W}{3} T] + \Lambda_t^2~, 
\end{equation}  
in which $m_{W, 0}$ is the central value of the $W$ mass.  We may linearize
this expression in $m_t$ about the value $\Lambda_t = 174$ GeV, finding
% Eq. (13)
\beq \label{eqn:mtst}
m_t = (174 + 210.8 S + 72.0 T)~{\rm GeV}/c^2~.
\eeq
The effect of including the observed top quark mass (Tevatron average
\cite{TEVEWWG}),
% Eq. (14)
\beq \label{eqn:mtexp}
m_t = 173.1 \pm 1.3~{\rm GeV}/c^2~,
\eeq
is shown in Fig.\ \ref{fig:mt}.
Although it would be inappropriate to exploit Eq.\ (\ref{eqn:mtst}) to derive
a new constraint in parameter space for $S$ and $T$ very far from zero, we
use it only for $S$ and $T$ very near zero, consistent with Eqs.\
(\ref{eq:Sht}) and (\ref{eq:Tht}), respecting the constraints provided by
$m_t$ and $m_W$.

% This is Figure 3

The linear relation (\ref{eqn:mtst}), when combined with the experimental value
(\ref{eqn:mtexp}), leads to a very narrow allowed band in $(S,T$) space which
is almost perpendicular to the semi-major ellipse in Fig.\ \ref{fig:APV},
thus acting to drastically reduce the allowed $(S,T)$ region.  One may then ask
what experiments are likely to provide the most incisive further
constraints on $S$ and $T$.  The answer is those that provide bands nearly
perpendicular to the semi-major ellipse in Fig.\ \ref{fig:mt}.  For an
observable $\bar x$ whose $(S,T)$ dependence is expressed as $\bar x = \bar x_0
+ A S + B T$, such an observable would have
% Eq. (15)
\beq \label{eqn:best}
210.8 A + 72.0 B = 0~~,~~~{\rm or}~~~B/A = -2.93~~~.
\eeq
These ratios for the observables $(M_W,~g_L^2,~g_R^2,~\Gamma_{ll}(Z),~\sin^2
\theta^{\rm eff})$ listed in Table \ref{tab:inp} are (--1.55, --2.44, 0.21,
--4.33, --0.71), respectively.  Of these, the observable $g_L^2(\nu {\rm NC})$
has the $B/A$ ratio closest to that in (\ref{eqn:best}).  A proposal to
improve the measurement of this observable (along with several others
connected with neutrino scattering) has been made in Ref.\ \cite{Adams:2008},
which we now discuss.

% This is Section V
\section{CONSTRAINTS FROM NEUTRINO SCATTERING
\label{sec:nu}}

For several decades, neutrino deep inelastic scattering ($\nu$DIS) has shed
tremendous light on the electroweak interactions, through the measurement of
the ratio of neutral-current and charged-current events.  The quantities
$g_L^2$ and $g_R^2$ in Table \ref{tab:inp} reflect the contribution to $S,T$
constraints of the most recent $\nu$DIS experiment performed by the NuTeV
Collaboration \cite{NuTeV}.  These coupling constants may be expressed as

\begin{subequations}
% Eq. (16a)
\begin{equation}
\label{eq:gL}
g_L^2 = (2 g_L^{\nu} g_L^u)^2 + (2 g_L^{\nu} g_L^d)^2
      = \rho^2 (\frac{1}{2} - \sin^2 \theta_W + \frac{5}{9} \sin^4 \theta_W)
\end{equation}
and
% Eq. (16b)
\begin{equation}
\label{eq:gR}
g_R^2 = (2 g_L^{\nu} g_R^u)^2 + (2 g_L^{\nu} g_R^d)^2
      = \rho^2 (\frac{5}{9} \sin^4 \theta_W).
\end{equation}  
\end{subequations}
Here we have used the parameter $\rho = 1 + \alpha T$, where the fine structure
constant $\alpha$ is taken to have its value of $\sim 1/128 = 0.0078$ at the
electroweak scale.  Linearizing these equations in $S$ and $T$, one obtains the
$S$ and $T$ dependence noted in Table \ref{tab:inp}.

% This is Figure 4
\begin{figure}
\includegraphics[width=5in]{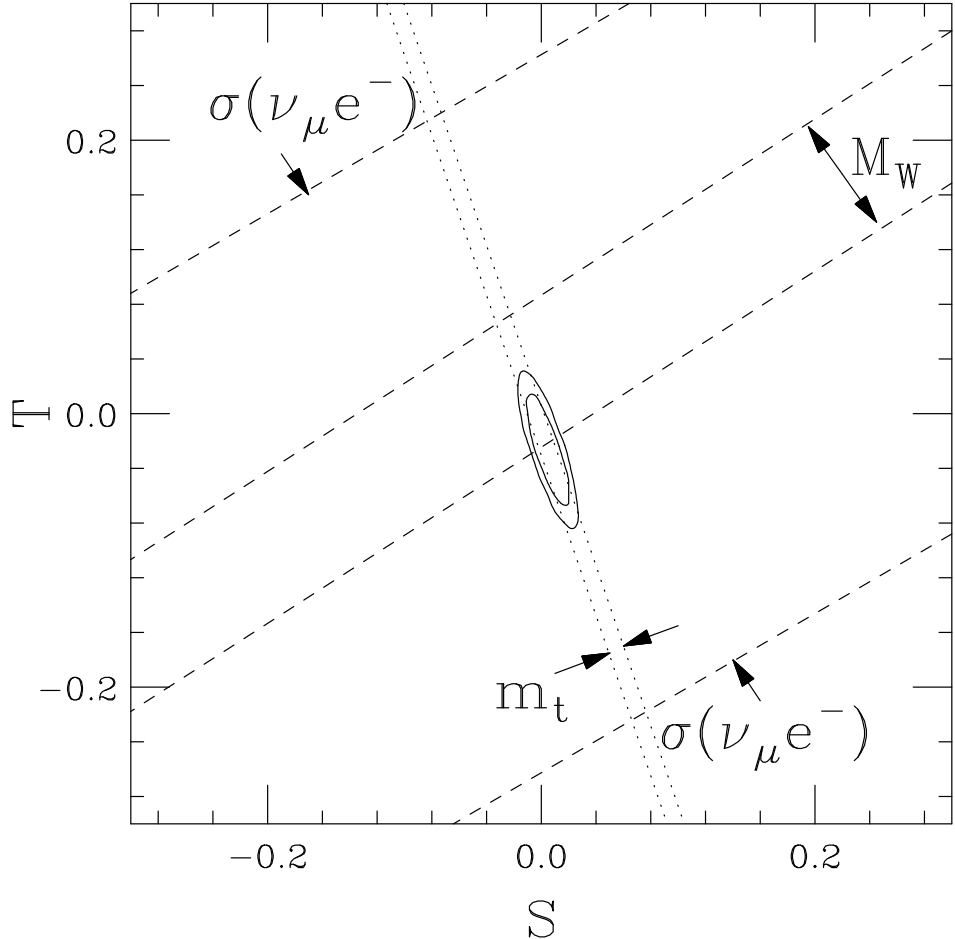}
\caption{Effects of including top quark mass constraint $m_t = 173.1 \pm 1.3$
GeV/$c^2$ on constraints in $S,T$ space.  The same data have been used as in
Fig.\ \ref{fig:mt} except that errors on $g_L^2$ and $g_R^2$ have been reduced
by a factor of 3, and constraints due to $\nu_\mu e^-$ elastic scattering (ES)
(broad dashed band) have been imposed.  Inner and outer ellipses denote 68\%
and 90\% confidence level (c.l.) limits.  Narrow dashed and dotted bands denote
constraints due to $M_W$ and $m_t$, respectively.}
\label{fig:mt2}
\end{figure}

The NuSOnG Collaboration has recently proposed remeasuring $g_L^2$ and
$g_R^2$ \cite{Adams:2008} with approximately a factor of two greater precision.
Assuming such errors and maintaining the same central values as reported by
NuTeV \cite{NuTeV}, we find hardly any effect on the allowed $S,T$ ellipse.
Assuming instead the errors on these quantities are divided by a factor of 3,
the result is shown in Fig.\ \ref{fig:mt2}.  In this Figure we have also
included the effects of measuring $\sigma(\nu_\mu e^- \to \nu_\mu e^-)$ to
0.7\% as advocated in Ref.\ \cite{Adams:2008}.  The inputs to the fit are
summarized in Table \ref{tab:inp2}.

% This is Table II
\begin{table}
\caption{Electroweak observables leading to $S,T$ constraints, including top
quark mass and possible improvements in neutrino scattering measurements.
\label{tab:inp2}}
\begin{center}
\begin{tabular}{l c c} \hline \hline
 & Experimental & Theoretical \\
Quantity & value & value \\ \hline
$Q_W({\rm Cs})$ & $-73.20 \pm 0.35$ & $-73.15 - 0.800S - 0.007T$ \\
$M_W({\rm GeV}/c^2)$ & $80.399 \pm 0.023$ & $80.385 - 0.29 S + 0.45 T$ \\
$g_L^2(\nu {\rm NC})$ \cite{NuTeV}$^a$ & $0.3027 \pm 0.0006$ & $0.3041
 -0.00272 S + 0.00665 T$ \\
$g_R^2(\nu {\rm NC})$ \cite{NuTeV}$^a$ & $0.03076\pm 0.0004$ & $0.0300
 + 0.00094 S - 0.00020 T$ \\
$\Gamma_{ll}(Z)$ (MeV) & $83.984 \pm 0.087$ & $84.005 - 0.18 S + 0.78 T$ \\
sin$^2 \theta^{\rm eff}$ & $0.23149 \pm 0.00016$ & $0.23140 + 0.00361 S
 - 0.00257 T$ \\
$m_t$ & $173.1 \pm 1.3$ & $174 + 210.8 S + 72.0 T$ \\
$\sigma(\nu_\mu) e^{-~b}$ & $0.356\pm0.0025$ & $0.356-0.00554 S + 0.00952 T$ \\
\hline \hline
\end{tabular}
\end{center}
\leftline{$^a$ Present errors assumed to be divided by 3.}
\leftline{$^b$ Anticipated \cite{Adams:2008}; units of $G_F^2 m_e E_\nu/(2
\pi)$.}
\end{table}

The $B/A$ ratio for $\nu_\mu e^-$ ES is --1.72, which would be quite favorable
for constraining the $(S,T)$ error ellipse if the uncertainty were
about 1/4 the value anticipated in Ref.\ \cite{Adams:2008}.

\section{CONCLUSIONS AND OUTLOOK
\label{sec:con}}

The $(S,T)$ language is a broad signal for BSM physics.  A complete set
of observables of the SM are expressible to first order in terms of $S, T$ and,
resultantly, separate measurements of SM observables may independently and
uniquely constrain the space of OCs.  It then is possible to quickly evaluate
the impact of any improvements in present experiments or totally new
measurements.

In spite of the sensitivity of atomic parity violation measurements to $S$,
we have seen that order-of-magnitude reductions to $Q_W$ are likely necessary
(compared to the $\sim 50 \%$ reductions that represent improvements in the
past eight years) before such experiments significantly constrain $S$.

Similarly ambitious undertakings in neutrino physics will be needed in
order to provide significant improvements in $(S,T)$ constraints.  Both
measurements of $\nu$DIS and $\nu_\mu e^-$ ES, foreseen in Ref.\
\cite{Adams:2008}, will have to represent ambitious, order-of-magnitude
improvements over current experiments in order to have a significant
impact on the $(S,T)$ plane.

\section*{Acknowledgements}

We thank Jens Erler and Paul Langacker for helpful communications.
T.~H. thanks the Enrico Fermi Institute and University of Chicago
Department of Physics for their hospitality.  This work was supported in
part by the U. S. Department of Energy through Grant No.\ DE-FG02-90ER40560.

%%%%%%%%%%%%%%%%%%%%%%%%%%%%%%%%%%%%%%%%%%%%%%%%%%%%%%%%%%%%%%%%%%%%%%%%%


\begin{thebibliography}{99}

\bibitem{PT}
M.~E.~Peskin and T.~Takeuchi,
Phys.\ Rev.\ Lett.\ {\bf 65}, 964 (1990); Phys.\ Rev.\ D {\bf 46}, 381 (1992).

\bibitem{MR}
W.~J.~Marciano and J.~L.~Rosner, Phys.\ Rev.\ Lett.\ {\bf 65}, 2963 (1990);
Erratum, {\it ibid.} {\bf 68}, 898 (1992).

\bibitem{Rosner:2001ck}
J. L. Rosner, Phys.\ Rev.\ D {\bf 65}, 073026 (2002). 

\bibitem{Sandars} P. G. H. Sandars, J. Phys.\ B {\bf B 23}, L655 (1990).

\bibitem{Rosner:2004}
J.~L.~Rosner, Phys.\ Rev.\ D {\bf 70}, 037301 (2004).
%%CITATION = PHRVA,D70,037301;%%

\bibitem{Wood:1997}
C. S. Wood {\it et al.}, Science {\bf 275}, 1759 (1997).

\bibitem{Guena:2005}
J. Gu\'ena, M. Lintz, and M. A. Bouchiat, Phys.\ Rev.\ A {\bf 71}, 042108
(2005).

\bibitem{Porsev:2009pr}
S.~G.~Porsev, K.~Beloy, and A.~Derevianko, Phys.\ Rev.\ Lett.\ {\bf 102},
181601 (2009).

\bibitem{Erler} J. Erler, A. Kurylov, and M. J. Ramsey-Musolf, Phys.\ Rev.\
D {\bf 68}, 016006 (2003); J. Erler and M. J. Ramsey-Musolf, Phys.\ Rev.\
D {\bf 72}, 073003 (2005).

\bibitem{Amsler:2008} C. Amsler {\it et al.} (Particle Data Group), Phys.\
Lett.\ B {\bf 667}, 1 (2008).

\bibitem{EL} J. Erler and P. Langacker, in K. Nakamura {\it et al.} (Particle
Data Group), J. Phys.\ G {\bf 37}, 075021 (2010), pp.\ 126--145.  See also
J. Erler and P. Langacker, arXiv:1003.3211v2, [Phys.\ Rev.\ Lett.\ (to be
published)].

\bibitem{NuTeV} G. P. Zeller {\it et al.} (NuTeV Collaboration), Phys.\ Rev.\
Lett.\ {\bf 88}, 091802 (2002); {\bf 90}, 239902(E) (2003).

\bibitem{LEPEWWG}
LEP Electroweak Working Group; see web page http://lepewwg.web.cern.ch/LEPEWWG
for periodic updates.  Our values are taken from Report No.\
CERN-PH-EP/2009-023, November 13, 2009.

\bibitem{TEVEWWG}
Tevatron Electroweak Working Group, Fermilab Report Fermilab-TM-2427-E,
arXiv:0903.2503.

\bibitem{Adams:2008}
T.~Adams, {\it et al.} (NuSOnG Collaboration), arXiv:0803.0354 [hep-ph] (2008),
Int.\ J.\ Mod.\ Phys.\ A {\bf 24}, 671 (2009).

\end{thebibliography}
\end{document}